\documentclass[conference]{IEEEtran}
\IEEEoverridecommandlockouts
\usepackage{cite}
\usepackage{amsmath,amssymb,amsfonts}
\usepackage{graphicx}
\usepackage{textcomp}
\usepackage{xcolor}
\usepackage{algpseudocode}
\usepackage{algorithm}
\def\BibTeX{{\rm B\kern-.05em{\sc i\kern-.025em b}\kern-.08em
    T\kern-.1667em\lower.7ex\hbox{E}\kern-.125emX}}
    
\usepackage{amsmath,amsthm}
\usepackage{color}
\usepackage{bbm,bm}
\usepackage{pgfplots}

\usepackage{tikz}
\usetikzlibrary{matrix,positioning,fit,calc,shapes}
\pgfplotsset{compat=newest}
\newcommand{\Var}{\mathrm{Var}}

\usepackage{graphicx}
\usepackage{color}
\usepackage{epsfig, amsmath, amsthm, amssymb,latexsym,amsmath,amsbsy, rotating}
\usepackage{amsfonts}
\usepackage{tikz}

\newtheorem{theorem}{Theorem}
\newtheorem{lemma}[theorem]{Lemma}
\newtheorem{corollary}[theorem]{Corollary}

\newtheorem{remark}{Remark}

\theoremstyle{definition}

\newcommand{\CLASSINPUTtoptextmargin}{2cm}

\usepackage{hyperref}
\usepackage{xcolor}
\hypersetup{
    colorlinks,
    linkcolor={black},
    citecolor={black},
    urlcolor={black}
}

\DeclareFontFamily{U}{futm}{}
\DeclareFontShape{U}{futm}{m}{n}{
  <-> s * [.92] fourier-bb
  }{}
\DeclareSymbolFont{Ufutm}{U}{futm}{m}{n}
\DeclareSymbolFontAlphabet{\mathbb}{Ufutm}

\begin{document}

\title{Shannon Bounds on Lossy Gray-Wyner Networks}


\author{\IEEEauthorblockN{Erixhen Sula}
\IEEEauthorblockA{\textit{The Department of Electrical Engineering and Computer Science} \\
\textit{MIT}\\
Massachusetts, USA \\
esula@mit.edu}
\and
\IEEEauthorblockN{Michael Gastpar}
\IEEEauthorblockA{\textit{School of Computer and Communication Sciences} \\
\textit{EPFL}\\
Lausanne, Switzerland \\
michael.gastpar@epfl.ch}
}

\maketitle

\begin{abstract}
The Gray-Wyner network subject to a fidelity criterion is studied. Upper and lower bounds for the trade-offs between the private sum-rate and the common rate are obtained for arbitrary sources subject to mean-squared error distortion. The bounds meet exactly, leading to the computation of the rate region, when the source is jointly Gaussian. They meet partially when the sources are modeled via an additive Gaussian ``channel". The bounds are inspired from the Shannon bounds on the rate-distortion problem.
\end{abstract}

\IEEEpeerreviewmaketitle

\section{Introduction}
Source coding in information theory establishes the limits of data compression. 
Various network source coding problems have been studied, most notably the problem of Slepian and Wolf~\cite{Slepian--Wolf} and the problem of Gray and Wyner~\cite{Gray--Wyner74}.
The latter is a broadcast situation composed of an encoder, which encodes the correlated sources, $(X,Y),$ into a common message and two private messages and two decoders that want to recover the respective source (either $X$ or $Y$), based on the common message and respective private message received.
Both in the lossless and lossy case, Gray and Wyner fully characterized the rate regions in~\cite{Gray--Wyner74}, up to the optimization over an auxiliary random variable. For jointly Gaussian sources, the rate region is partially computed in \cite{Xu--Liu--Chen,Akyol--Rose--2014} and later on the rate region is computed exactly in \cite{Sula-Gastpar22}. For any other sources the rate region remains unknown. Other related work include \cite{Zheng--Xu--2020,Lapidoth--Wigger,Kumar--Li--Gamal,Veld--Gastpar}.
In the present paper:
\begin{itemize}
\item We provide upper and lower bounds for the private sum-rate versus the common rate of the lossy Gray-Wyner network under mean-squared error distortion for any arbitrary source, akin to the bounds established by Shannon for rate distortion problem.
\item As a special case when the source is jointly Gaussian the upper and lower bounds meet exactly.
\item We further improve upper bounds for a class of sources referred to as additive Gaussian ``channel" sources, leading to a partial computation of the private sum-rate versus the common rate region of the lossy Gray-Wyner network.
\end{itemize}

\subsection{Organization}
In Section \ref{sec:probstate}, we describe the problem of the lossy Gray-Wyner network. In Section \ref{sec:Shannonbounds}, we briefly review the Shannon bounds for the rate-distortion problem and further extend the bounds for the conditional rate-distortion problem. In Section \ref{sec:mainresult}, we state the main results of the paper, that is the upper and the lower bound for the private sum-rate versus the common rate region of the lossy Gray-Wyner network, in a similar fashion as the Shannon bounds for the rate-distortion problem. In Section \ref{sec:jointlyGauss}, when the sources are jointly Gaussian as a special case of our main result we compute the rate region of the lossy Gray-Wyner network. In Section \ref{sec:additiveGauss}, for additive Gaussian ``channel" sources, we partially compute the rate region of the lossy Gray-Wyner network.


\section{Problem Statement} \label{sec:probstate}
\begin{figure}[h!] 
\centering
\begin{tikzpicture}[scale=0.8]
\draw[black,thick] (0,1) rectangle (1,2);
\draw[black,thick,anchor=east] (-0.5,1.5) node{$(X,Y)$};
\draw[black,thick,anchor=west] (5.5,3) node{$\hat{X}$};
\draw[black,thick,anchor=west] (5.5,0) node{$\hat{Y}$};
\draw[black,thick,anchor=center] (0.5,1.5) node{$\mathcal{E}$};
\draw[black,thick,anchor=south] (2,0) node{$R_y$};
\draw[black,thick,anchor=south] (2,1.5) node{$R_c$};
\draw[black,thick,anchor=south] (2,3) node{$R_x$};
\draw[black,thick](-0.5,1.5)--(0,1.5);
\draw[->,black,thick](0.5,1)--(0.5,0)--(3.5,0);
\draw[->,black,thick](1,1.5)--(4,1.5)--(4,2.5);
\draw[->,black,thick](4,1.5)--(4,0.5);
\draw[->,black,thick](0.5,2)--(0.5,3)--(3.5,3);
\draw[black,thick,anchor=center] (4,0) node{$\mathcal{D}_y$};
\draw[black,thick,anchor=center] (4,3) node{$\mathcal{D}_x$};
\draw[black,thick] (3.5,-0.5) rectangle (4.5,0.5);
\draw[black,thick] (3.5,2.5) rectangle (4.5,3.5);
\draw[->,black,thick](4.5,0)--(5.5,0);
\draw[->,black,thick](4.5,3)--(5.5,3);
\end{tikzpicture}
\caption{The Gray-Wyner Network} 
\label{fig:Gray-Wyner}
\end{figure}
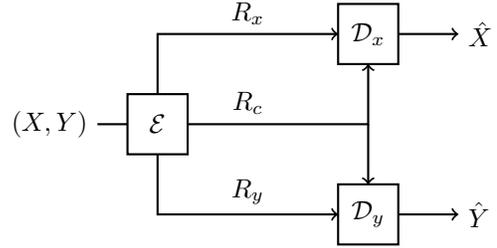

Let the correlated source $(X,Y)$ have a joint distribution that take values in $\mathcal{X} \times \mathcal{Y}$. Let $S_c,S_x$ and $S_y$ be messages represented by $nR_c,nR_x$ and $nR_y$ bits. We use a block code of length $n$ and the encoder/decoder system is given by
\begin{align}
(S_c,S_x,S_y)=f_{\mathcal{E}}(X^n,Y^n),
\end{align} 
where $f_{\mathcal{E}}(.)$ is the encoding function, $X^n \in \mathcal{X}^n$ and $Y^n \in \mathcal{Y}^n$ and 
\begin{align}
\hat{X}^n=f_{\mathcal{D}_x}(S_c,S_x), \quad \hat{Y}^n=f_{\mathcal{D}_y}(S_c,S_y),
\end{align} 
where $f_{\mathcal{D}_x}(.),f_{\mathcal{D}_y}(.)$ are the decoding function, $\hat{X}^n \in \hat{\mathcal{X}}^n$ and $\hat{Y}^n \in \hat{\mathcal{Y}}^n$. The system has an error rate of $(P_{e_x},P_{e_y})$, where
\begin{align}
P_{e_x}=\mathbb{E}\left[ \frac{1}{n} \sum_{k=1}^n d_x(X_k,\hat{X}_k) \right],  P_{e_y}=\mathbb{E}\left[ \frac{1}{n} \sum_{k=1}^n d_y(Y_k,\hat{Y}_k) \right].
\end{align}
Let the event $\mathcal{E}_r$ be 
\begin{align}
\mathcal{E}_r=\left\{ \Delta_x <P_{e_x} \right\} \cup \left\{ \Delta_y <P_{e_y} \right\}.
\end{align}

A rate triple $(R_c,R_x,R_y)$ is said to be $(\Delta_x,\Delta_y)$-achievable if, for any specified positive error probability $P_e$ and sufficiently large $n$, there are encoding and decoding functions that satisfy $\mathrm{Pr}[\mathcal{E}_r] \leq P_e.$ The closure of the set of achievable $(R_c,R_x,R_y)$ is called $\mathcal{R}(\Delta_x,\Delta_y)$.

The main result of~\cite[Theorem 6-8]{Gray--Wyner74}, says that the rate region is given by the closure of the union of the regions
\begin{align} \label{eqn:main}
\mathcal{R}(\Delta_x,\Delta_y)=\{ &(R_{\sc c},R_x,R_y):R_c \ge I(X,Y;W),   \\
&R_x \ge R_{X|W}(\Delta_x) , R_y \ge R_{Y|W}(\Delta_y) \},\label{eq-GrayWyner}
\end{align}
where the union is over all probability distributions $p(w, x, y)$ with marginals $p(x,y)$ and here $R_{X|W}$ and $R_{Y |W}$ are conditional rate-distortion function \cite{Gray--1972}.

Instead of considering the triple $(R_c,R_x,R_y)$ we focus on the common rate versus sum of private rates tradeoff $(R_c,R_x+R_y)$ for symmetric distortion $\Delta=\Delta_x=\Delta_y$, that motivates the following object
\begin{align} \label{eqn:GWdef}
R_c(R_p)=\inf_{W:R_{X|W}(\Delta)+R_{Y|W}(\Delta)\leq R_p} I(X,Y;W).
\end{align}


%
\section{Shannon Bounds on the Rate-Distortion} \label{sec:Shannonbounds}
The bounds on the rate-distortion function are 
\begin{align}
\frac{1}{2}\log^+{\frac{N(X)}{\Delta}} \leq R_X(\Delta) \leq \frac{1}{2}\log^+{\frac{\Var(X)}{\Delta}},
\end{align}
where $N(X):=\frac{1}{2 \pi e} e^{2h(X)}$ and the rate distortion function is defined as
$R_{X}(\Delta):=\inf_{\hat{X}:\mathbb{E}[(X-\hat{X})^2]\leq \Delta} I(X;\hat{X}).$
We extend the result for conditional random variables as follows:
\begin{lemma} \label{lem:cond-rate-distor}
The bounds on the \emph{conditional} rate-distortion function are
\begin{align}
\frac{1}{2}\log^+{\frac{N(X|Y)}{\Delta}} \leq R_{X|Y}(\Delta) \leq \frac{1}{2}\log^+{\frac{\mathbb{E}[\Var(X|Y)]}{\Delta}},
\end{align}
where $N(X|Y):=\frac{1}{2 \pi e} e^{2h(X|Y)}$ and the rate distortion function is defined as 
$R_{X|Y}(\Delta):=\inf_{\hat{X}:\mathbb{E}[(X-\hat{X})^2]\leq \Delta} I(X;\hat{X}|Y).$
\end{lemma}
Proof is given in Appendix \ref{app:proofcondratedistBB}. To the best of our knowledge, the upper bound we derived on conditional rate distortion problem do not appear in the literature.

\section{Main Result} \label{sec:mainresult}
\begin{theorem} \label{thm:main}
Let $(X,Y)$ be an arbitrary source with unit variance. Let the distortion $d_x(.)$ and $d_y(.)$ be mean-squared error measure. Then,
\begin{align} \label{eqn:partoneLUB}
&\frac{1}{2} \log^{+}{\frac{N^2(X,Y)}{(1-\rho) \left( 2\Delta e^{R_p}+\rho-1 \right)}} \leq R_c(R_p) \nonumber \\
& \hspace{8em} \leq \frac{1}{2} \log^{+}{\frac{1-\rho^2}{(1-\rho) \left( 2\Delta e^{R_p}+\rho-1 \right)}}
\end{align}
for $1-\rho \le \Delta e^{R_p} \le 1$ and
\begin{align} \label{eqn:parttwoLUB}
 \frac{1}{2} \log^{+}{\frac{N^2(X,Y)}{\Delta^2 e^{2R_p}}} \leq R_c(R_p) \leq \frac{1}{2} \log^{+}{\frac{1-\rho^2}{\Delta^2 e^{2R_p}}} 
\end{align}
for $\Delta e^{R_p} \le 1-\rho$, where $N(X,Y)$ denotes the entropy power of the pair $(X,Y)$, $N(X,Y):=\frac{1}{2 \pi e} e^{h(X,Y)}.$
\end{theorem} 
Proof is given in Appendix \ref{app:proofmainthmLB} and \ref{app:proofmainthmUB}.
\begin{remark} \label{rem:unitvariance}
For sources $X$ and $Y$ of variance $\sigma^2,$ it suffices to replace $\Delta$ with $\Delta/{\sigma}^2,$ and the entropy power is still computed as if variances were unity.
\end{remark}
\begin{remark}
Note that $N(X,Y)\le \sqrt{1-\rho^2},$ with equality if and only if the pair $(X,Y)$ is jointly Gaussian.
\end{remark}

\section{Jointly Gaussian Source} \label{sec:jointlyGauss}
When $(X,Y)$ is jointly Gaussian where the variance of $X$ and $Y$ is one and the correlation coefficient is $\rho$, then $N(X,Y)=\sqrt{1-\rho^2}$, thus the upper and the lower bound in Theorem \ref{thm:main} meet with equality, thus we recover the result in \cite[Theorem~1]{Sula-Gastpar22} and \cite[Theorem~1]{Sula-Gastpar20CISS}.
\begin{corollary}[Theorem 1 in \cite{Sula-Gastpar22}] \label{cor:Gauss}
Let $(X,Y)$ be jointly Gaussian with correlation coefficient $\rho$ and unit variance. Let the distortion $d_x(.)$ and $d_y(.)$ be mean-squared error measure then,
\begin{align} \label{eqn:partoneLUBGauss}
 R_c(R_p) = \frac{1}{2} \log^{+}{\frac{1-\rho^2}{(1-\rho) \left( 2\Delta e^{R_p}+\rho-1 \right)}}
\end{align}
for $1-\rho \le \Delta e^{R_p} \le 1$ and
\begin{align} \label{eqn:parttwoLUBGauss}
R_c(R_p) = \frac{1}{2} \log^{+}{\frac{1-\rho^2}{\Delta^2 e^{2R_p}}} 
\end{align}
for $\Delta e^{R_p} \le 1-\rho$.
\end{corollary}

\section{Additive Gaussian ``Channel'' Source} \label{sec:additiveGauss}
Let us consider the lossy Gray-Wyner network where the source is modeled as follows
\begin{align}
X&=\theta +Z_x, \nonumber \\
Y&=\theta +Z_y, \label{eqn:Gaussianmodel}
\end{align}
where $X$ and $Y$ have variance one and correlation coefficient $\rho$, $\theta$ that is mean zero and variance $\sigma^2_{\theta}\le 1$, is independent of $(Z_x,Z_y) \sim \mathcal{N}(0,K_{(Z_x,Z_y)})$ and 
\begin{align}
K_{(Z_x,Z_y)}=\begin{pmatrix}
1-\sigma^2_{\theta} & \rho-\sigma^2_{\theta} \\
\rho-\sigma^2_{\theta} & 1-\sigma^2_{\theta}
\end{pmatrix}.
\end{align}
Refer to Remark \ref{rem:unitvariance} for sources $X$ and $Y$ with variance different than one.

\begin{theorem} \label{thm:addGauss}
For the additive Gaussian source $(X,Y)$ described in (\ref{eqn:Gaussianmodel}), we compute the exact trade-offs between the private sum-rate and the common rate
\begin{align} \label{eqn:addGaussp-one}
R_c(R_p) = \frac{1}{2} \log^{+}{\frac{N^2(X,Y)}{(1-\rho) \left( 2\Delta e^{R_p}+\rho-1 \right)}},
\end{align}
for $1-\rho \le \Delta e^{R_p} \le 1 -\sigma^2_{\theta}$ and 
\begin{align} \label{eqn:addGaussp-two}
R_c(R_p) = \frac{1}{2} \log^{+}{\frac{N^2(X,Y)}{\Delta^2 e^{2R_p}}},
\end{align}
for $\Delta e^{R_p}  \le  1-\rho$.
\end{theorem}

Proof is given in Appendix \ref{app:proofaddGauss}. In Figure \ref{fig:AdditiveGauss}, we plot schematically the sum of private rates $R_p$ versus the common rate $R_c$. In \cite{Yang-Chen14}, for the model in (\ref{eqn:Gaussianmodel}), they computed Wyner's common information 
\begin{align}
C(X;Y)=\log{\frac{N(X,Y)}{1-\rho}},
\end{align}
where Wyner's common information is defined as $C(X;Y):= \inf\limits_{W:X-W-Y} I(X,Y;W)$. The computation of Wyner's common information corresponds to the rate pair 
\begin{align}
(R_c,R_p)=\left(\log{\frac{N(X,Y)}{1-\rho}},\log{\frac{1-\rho}{\Delta}} \right),
\end{align}
that is shown in Figure \ref{fig:AdditiveGauss}.
\begin{figure}[h!] 
\scalebox{0.82}{\centering{
\begin{tikzpicture}
\draw[->, line width=1pt](0,0)--(8,0);
\draw[black,thick,anchor=center] (6.75,0.2) node{$?$};
\draw[dashed, line width=1pt](0,2)--(2,2)--(2,0);
\draw[dashed, line width=1pt](2,2)--(5,-0.7);
\draw[black,thick,anchor=north] (5,-0.7) node{$R_c(R_p)+R_p=\log{\frac{N(X,Y)}{\Delta}}$};
\draw[->, line width=1pt](0,0)--(0,5);
\draw[black,thick,anchor=north] (8.3,0) node{$R_p$};
\draw[black,thick,anchor=north] (6,0) node{$\log{\frac{1-\sigma^2_{\theta}}{\Delta}}$};
\draw[black,thick,anchor=north] (2,0) node{$\log{\frac{1-\rho}{\Delta}}$};
\draw[black,thick,anchor=east] (0,4) node{$\log{\frac{N(X,Y)}{\Delta}}$};
\draw[black,thick,anchor=east] (0,0.25) node{$I(\theta;X,Y)$};
\draw[black,thick,anchor=center] (0,5.5) node{$R_c(R_p)$};
\draw[line width=1pt](0,4)--(2,2);
\draw[dashed, line width=1pt](6,0)--(6,0.25);
\draw[dashed, line width=1pt](0,0.25)--(6,0.25);
\draw[black,thick,anchor=east] (0,2) node{$\log{\frac{N(X,Y)}{1-\rho}}$};
\draw[black,thick, line width=1pt] (2,2) to [out =322, in =168] (6,0.25);
\end{tikzpicture}}}
\caption{Private sum-rate versus common rate for lossy Gray-Wyner network when source $(X,Y)$ is modeled in (\ref{eqn:Gaussianmodel}) as additive Gaussian ``channels".}
\label{fig:AdditiveGauss}
\end{figure}
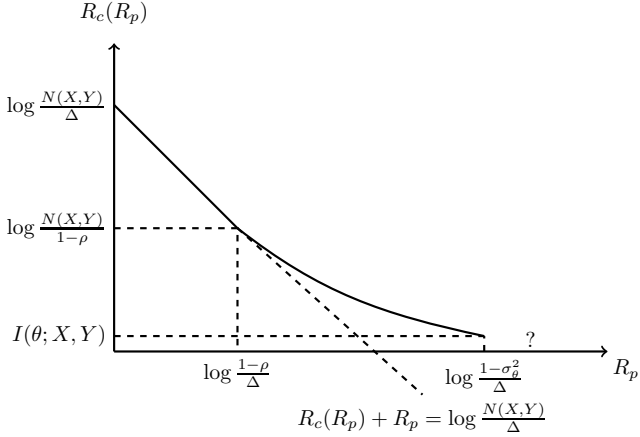



\appendices
\section{Proof of Lemma \ref{lem:cond-rate-distor}} \label{app:proofcondratedistBB}
The lower bound is derived below  \begin{align}
R_{X|W}(\Delta) &\geq  R_{X}(\Delta)-I(X;W) \label{eqn:Gray73} \\
& \geq \frac{1}{2} \log{\frac{N(X)}{\Delta}} -I(X;W) \label{eqn:ShannonLB} \\
& = \frac{1}{2} \log{\frac{N(X|W)}{\Delta}} \label{eqn:defEntpower}
\end{align}
where (\ref{eqn:Gray73}) is a consequence of \cite[Theorem~3.1]{Gray73}, (\ref{eqn:ShannonLB}) follows from the Shannon's lower bound on rate-distortion and (\ref{eqn:defEntpower}) follows from the definition of the entropy power and conditional entropy power. 
%
%
%

To derive the upper bound, we start by constructing $\hat{X}$ and we firstly will show that
\begin{align} \label{eqn:conditionalratedistortionshannon}
R_{X|W}(\Delta) \leq \mathbb{E} \left[ \frac{1}{2}\log^+{\frac{\Var(X|W)}{\Delta_W}} \right],
\end{align}
where $\Delta=\mathbb{E}[\Delta_W]$. For simplicity, we consider $W$ to be discrete. Let $Z_w\sim \mathcal{N}\left(0, \Var(Z_w) \right)$ be independent of $X_w$ ($X_w:=X|W=w$) for any instance $w$ 
\begin{align} \label{eqn:relvarzwxw}
\Var(Z_w)=\frac{\Var(X_w)\Delta_w}{\Var(X_w)-\Delta_w},
\end{align}
and let us construct
\begin{align}  \label{eqn:defXw}
\hat{X}_w=\alpha(X_w+Z_w),
\end{align}
where 
\begin{align} \label{eqn:defalpha}
\alpha=\frac{\Var(X_w)-\Delta_w}{\Var(X_w)}.
\end{align}
Without loss of optimality we can assume $X_w$ to be zero mean because $\sum_w p(w)I(X_w;\hat{X}_w)$ is not effected by the mean and $\sum_{w}p(w) \mathbb{E}[(X_w-\hat{X}_w)^2]$ does not change if $\mathbb{E}[\hat{X}_w]=\mathbb{E}[X_w]$. From (\ref{eqn:defXw}) $\mathbb{E}[\hat{X}_w]=\mathbb{E}[X_w]$ is satisfied if we choose $\mathbb{E}[Z_w]=\frac{1-\alpha}{\alpha}\mathbb{E}[X_w]$, which means that neither the objective nor the constraint of the problem changes. Therefore, we will assume that $X_w$ and $Z_w$ have mean zero. Thus,
\begin{align}
\mathbb{E}[(X-\hat{X})^2]&=\sum_{w}p(w) \mathbb{E}[(X-\hat{X})^2|W=w] \\
&=\sum_{w}p(w) \mathbb{E}[(X_w-\hat{X}_w)^2] \\
&=\sum_{w}p(w) \left( (1-\alpha)^2 \Var(X_w) + \alpha^2 \Var(Z_w) \right)   \label{eqn:meanXequalmeanXhat} \\
&= \sum_w p(w) \Delta_w  \label{eqn:subalpha} \\
&=\Delta \label{eqn:defdelta}
\end{align}
where (\ref{eqn:meanXequalmeanXhat}) follows from (\ref{eqn:defXw}). Equation (\ref{eqn:subalpha}) follows by (\ref{eqn:relvarzwxw}) and (\ref{eqn:defalpha}). Equation (\ref{eqn:defdelta}) is the definition of $\Delta$. On the other hand, we compute
\begin{align}
I(X;\hat{X}|W=w)&=h(\hat{X}_w)-h(\hat{X}_w|X_w) \\
&=h(\hat{X}_w)-h\left(\alpha Z_w\right) \label{eqn:condratedistachievable1} \\
&=h(X_w+Z_w)-h\left(Z_w\right) \label{eqn:condratedistachievable2} \\
&\leq \frac{1}{2}\log\frac{\Var(X_w+Z_w)}{\Var(Z_w)} \label{eqn:condratedistachievable3} \\
&= \frac{1}{2}\log\frac{\Var(X_w)+\Var(Z_w)}{\Var(Z_w)} \label{eqn:condratedistachievable4} \\
&= \frac{1}{2}\log\frac{\Var(X_w)}{\Delta_w}. \label{eqn:condratedistachievable5}
\end{align}
where (\ref{eqn:condratedistachievable1}) follows by dropping the information about $X$ and $X_w$ is independent of $Z_w$; (\ref{eqn:condratedistachievable2}) follows from the property $h(\alpha X)=\log{|\alpha|}+ h(X)$; (\ref{eqn:condratedistachievable3}) follows from $h(X_w+Z_w) \leq \frac{1}{2}\log{\left(2 \pi e \Var(X_w+Z_w) \right)}$; (\ref{eqn:condratedistachievable4}) follows from independence of $X_w$ and $Z_w$; (\ref{eqn:condratedistachievable5}) follows from (\ref{eqn:relvarzwxw}). To finalize the proof we need to solve the distortion allocation problem in (\ref{eqn:conditionalratedistortionshannon}).
%
Let us assume $W$ has an alphabet size $|\mathcal{W}|=n$. Without loss of optimality we can assume that $\Var(X|W=w_1) \leq \Var(X|W=w_2) \leq \dots  \leq \Var(X|W=w_n)$.
Then, the optimal solution of (\ref{eqn:conditionalratedistortionshannon}) starts by setting $\gamma = \Delta$ and check if $\gamma \leq \Var(X|W=w_i)$, $i=1,\dots,n$. If condition is met, we stop and the optimal solution is $\Delta_{w_i}=\gamma$ for $i=1,\dots,n$. If condition is not met, we compute 
\begin{align}
\gamma= \frac{\Delta- p(w_1) \Var(X|W={w_1})}{1- p(w_1)},
\end{align}
and check if $\gamma \leq \Var(X|W=w_i)$ for $i=2,\dots,n$. If condition is met, we stop and the optimal solution is $\Delta_{w_1}=\Var(X|W=w_1)$ and $\Delta_{w_i}=\gamma$ for $i=2,\dots,n$. If condition is not met, we compute 
\begin{align}
\gamma= \frac{\Delta- \sum_{i=1}^2 p(w_i) \Var(X|W={w_i})}{1- \sum_{i=1}^2 p(w_i)}.
\end{align}
We repeat this procedure until it stops. Let us assume that the procedure stops for
\begin{align}
\gamma=\frac{\Delta-\sum_{i=1}^k p(w_i) \Var(X|W=w_i)}{1-\sum_{i=1}^k p(w_i)},
\end{align}
which means that $\gamma \leq \Var(X|W=w_i)$ for $i=k+1,\dots,n$. Thus the optimal solution is $\Delta_{w_i}=\Var(X|W=w_i)$ for $i=1,\dots,k$ and $\Delta_{w_i}=\gamma$ for $i=k+1,\dots,n$. Let us denote with 
\begin{align}
&x:=\frac{\sum_{i=1}^k p(w_i) \Var(X|W=w_i)}{\gamma} \label{eqn:defx}, \\
&y:=\frac{\sum_{i=k+1}^n p(w_i) \Var(X|W=w_i)}{\gamma} \label{eqn:defy}.
\end{align}
Thus, $x \leq \sum_{i=1}^k p(w_i)$ and $y \geq \sum_{i=k+1}^n p(w_i)$ which follows from the optimal solution $\Var(X|W=w_1)\leq \dots \leq \Var(X|W=w_k) \leq \gamma \leq \Var(X|W=w_{k+1}) \leq \dots \leq \Var(X|W=w_n)$. For simplicity let us denote with $\alpha_x:=\sum_{i=1}^k p(w_i)$ and $\alpha_y:=\sum_{i=k+1}^n p(w_i)$. Note that $\alpha_x+\alpha_y=1$. We further upper bound (\ref{eqn:conditionalratedistortionshannon}) as follows

\begin{align}
\mathbb{E} \left[ \frac{1}{2}\right. & \left.\log^+{\frac{\Var(X|W)}{\Delta_W}} \right] \\
&= \sum_{i=k+1}^n \frac{p(w_i)}{2} \log{\frac{\Var(X|W=w_i)}{\gamma}} \label{eqn:nfplogplus1} \\
& \leq \frac{1}{2} \log{ \left( \sum_{i=1}^k p(w_i) +\sum_{i=k+1}^n p(w_i) \frac{\Var(X|W=w_i)}{\gamma} \right)} \label{eqn:nfplogplus2} \\
& = \frac{1}{2} \log{ \left( \alpha_x+y \right)} \label{eqn:nfplogplus3} \\
& \leq  \frac{1}{2} \log{ \left( \frac{x+y}{x+\alpha_y} \right)} \label{eqn:nfplogplus4} \\
&=\frac{1}{2} \log{\frac{\mathbb{E}[\Var(X|W)]}{\Delta}}, \label{eqn:nfplogplus5}
\end{align}
where (\ref{eqn:nfplogplus1}) follows from the optimal solution $\Delta_{w_i}=\Var(X|W=w_i)$ for $i=1,\dots,k$ and $\Delta_{w_i}=\gamma$ for $i=k+1,\dots,n$ as explained above; (\ref{eqn:nfplogplus2}) follows from Jensen's inequality; (\ref{eqn:nfplogplus3}) follows from definition of $x$ and $y$ in (\ref{eqn:defx}) and (\ref{eqn:defy}); (\ref{eqn:nfplogplus4}) follows by rearranging $\alpha_x+y \leq \frac{x+y}{x+\alpha_y}$ into $(x-\alpha_x)(y-\alpha_y) \leq 0$ which holds for $x \leq \alpha_x$ and $y \geq \alpha_y$ and (\ref{eqn:nfplogplus5}) follows from plugging back $x$ and $y$. The argument extends for $W$ is a continuous random variable. 
 
\section{Proof of Theorem \ref{thm:main} Lower Bound} \label{app:proofmainthmLB}
Let us start from $R_c(R_p)$ that we defined in (\ref{eqn:GWdef}),
\begin{align}
&R_c( R_p)=\inf_{W:R_{X|W}(\Delta)+R_{Y|W}(\Delta)\leq R_p} I(X,Y;W) \\
& \geq \inf_{W:\frac{1}{2} \log{\frac{N(X|W)}{\Delta}}+ \frac{1}{2} \log{\frac{N(Y|W)}{\Delta}} \leq R_p} I(X,Y;W) \label{eqn:defcondratedist} \\
& \geq \inf_W I(X,Y;W) + \frac{\nu}{2} \log{\frac{N(X|W)}{\Delta}} \nonumber \\
& \quad \quad + \nu \left( \frac{1}{2} \log{\frac{N(Y|W)}{\Delta}} - R_p \right) \label{eqn:weakduality} \\
& =  h(X,Y) - \nu R_p - \nu \log(2 \pi e \Delta) \nonumber \\
& \quad \quad + \nu \cdot \inf_W h(X|W)+h(Y|W)- \frac{1}{\nu}h(X,Y|W) \label{eqn:minsplit} \\
& \geq h(X,Y) - \nu R_p -\nu \log{(2 \pi e \Delta)}  \nonumber \\
& \quad \quad + \nu \cdot \hspace{-2em} \min_{0 \preceq K^{\prime} \preceq \begin{pmatrix} 1 & \rho \\ \rho &1  \end{pmatrix}} h(X^{\prime})+h(Y^{\prime})- \frac{1}{\nu}h(X^{\prime},Y^{\prime}) \label{eqn:twoboundeval} \\
& \geq \frac{1}{2} \log{(2\pi e)^2N^2(X,Y)} - \nu R_p -\nu \log{(2 \pi e \Delta)} \nonumber \\
& \quad \quad + \frac{\nu}{2} \log{\frac{\nu^2}{2\nu-1}}  -\frac{1-\nu}{2} \log{(2\pi e)^2\frac{(1-\rho)^2}{2\nu-1}} \label{eqn:hypereval2} \\
& = \left\{ \begin{array}{lr} \frac{1}{2} \log^{+}{\frac{N^2(X,Y)}{(1-\rho) \left( 2\Delta e^{R_p}+\rho-1 \right)}}, &   \mbox{ if } 1-\rho \le \Delta e^{R_p} \le 1  \\
 \frac{1}{2} \log^{+}{\frac{N^2(X,Y)}{\Delta^2 e^{2R_p}}} , &   \mbox{ if }  \Delta e^{R_p} \le 1-\rho.
\end{array} \right. \label{eqn:lastGrayWyner}
\end{align}
where \eqref{eqn:defcondratedist} follows from relaxing the constraint by Lemma \ref{lem:cond-rate-distor}; \eqref{eqn:weakduality} follows from weak duality for $ \nu \geq 0$; \eqref{eqn:minsplit} follows from the definition of conditional entropy power;
\eqref{eqn:twoboundeval} follows from \cite[Theorem~8]{Sula-Gastpar22}
 where $\frac{1}{2} <\nu \leq1$; (\ref{eqn:hypereval2}) follows from \cite[Lemma~13]{Sula-Gastpar22}
  for $\nu \geq \frac{1}{1+\rho}$; and (\ref{eqn:lastGrayWyner}) follows from maximizing 
\begin{align}
\ell(\nu)&:=\frac{1}{2} \log{(2\pi e)^2N^2(X,Y)} - \nu R_p -\nu \log{(2 \pi e \Delta)} \nonumber \\
&\quad \quad  + \frac{\nu}{2} \log{\frac{\nu^2}{2\nu-1}}-\frac{1-\nu}{2} \log{(2\pi e)^2\frac{(1-\rho)^2}{2\nu-1}},
\end{align}
for $1 \geq \nu \geq \frac{1}{1+\rho}$ or in other words we need to solve
$\max_{1 \geq \nu \geq \frac{1}{1+\rho}} \ell(\nu)$. Note that the function $\ell$ is concave since
\begin{align}
\frac{\partial^2 \ell}{\partial \nu^2}=-\frac{1}{\nu(2\nu-1)}<0,
\end{align}
and by studying the monotonicity
\begin{align}
\frac{\partial \ell}{\partial \nu}=\log{\frac{\nu(1-\rho)}{(2\nu-1)\Delta e^{R_p}}},
\end{align}
its maximal value occurs when the derivative vanishes, that is, when $\nu_*=\frac{\Delta e^{R_p}}{2 \Delta e^{R_p}-1+\rho}.$ Substituting for the optimal $\nu_*$ we get
\begin{align}
R_c( R_p) &\geq \ell \left(\frac{\Delta e^{R_p}}{2\Delta e^{R_p}-1+\rho} \right) \\
&=\frac{1}{2} \log^+ \frac{N^2(X,Y)}{(1-\rho) \left( 2 \Delta e^{R_p}-1+\rho \right) },
\end{align}
for $1 \geq \nu_* \geq \frac{1}{1+\rho}$, which means the expression is valid for $1-\rho \leq \Delta e^{R_p} \leq 1$.
The other case is $ \Delta e^{R_p} \leq 1-\rho$. In this case note that $\nu(1-\rho) \geq \nu \Delta e^{R_p} \geq (2\nu-1)\Delta e^{R_p}$ for $\nu \leq 1$. This implies $\frac{\nu(1-\rho)}{(2\nu-1) \Delta e^{R_p}} \geq 1$, thus we have $\frac{\partial \ell}{\partial \nu} \geq 0$. Since the function is concave and increasing the maximum is attained at $\nu_*=1$, thus 
\begin{align}
R_c(R_p) \geq \ell \left(1 \right) =\frac{1}{2} \log^+ \frac{N^2(X,Y)}{\Delta^2 e^{2R_p}},
\end{align}
where the expression is valid for $\Delta e^{R_p} \leq 1-\rho$.
The exact formula is derived assuming unit-variance sources.

\section{Proof of Theorem \ref{thm:main} Upper Bound} \label{app:proofmainthmUB}

\subsection{$W$ is a random vector of size one}
We managed to show in Lemma \ref{lem:cond-rate-distor} that $R_{X|W}(\Delta) \leq \frac{1}{2} \log{\frac{\mathbb{E}[\Var(X|W)]}{\Delta}}$ and in order to satisfy the constraint in (\ref{eqn:GWdef}) we need to finally show that $\mathbb{E}[\Var(X|W)] \leq \Delta e^{R_p}$. We construct $W$ as
\begin{align}
W=\alpha(X+Y) + N,
\end{align}
where $N$ is independent of $(X,Y)$ and $N\sim \mathcal{N}\left(0,\frac{2\Delta e^{R_p}+\rho -1}{1+\rho}\right)$ and we choose $\alpha=\frac{\sqrt{1-\Delta e^{R_p}}}{1+\rho}$, thus

\begin{align}
\mathbb{E}[\Var(X|W)]&= \Var(X)-\Var(\mathbb{E}[X|W]) \\
&= \mathbb{E}[X^2] - \mathbb{E}[\mathbb{E}^2[X|W]] \\
& \leq \mathbb{E}[X^2] - \frac{\mathbb{E}^2[W\mathbb{E}[X|W]]}{\mathbb{E}[W^2]} \label{eqn:appchauchy} \\
&=\frac{\mathbb{E}[X^2]\mathbb{E}[W^2]-\mathbb{E}^2[XW]}{\mathbb{E}[W^2]}\\
&=\Delta e^{R_p} \label{eqn:subalphasqrt}
\end{align}
where (\ref{eqn:appchauchy}) follows from Cauchy–Schwarz inequality and (\ref{eqn:subalphasqrt}) follows from $\mathbb{E}[XW]=\sqrt{1-\Delta e^{R_p}}$. Thus,
\begin{align}
R_c(R_p)\leq I(X,Y;W)&=h(W)-h(N) \\
&\leq \frac{1}{2}\log{\frac{\Var(W)}{\Var(N)}} \label{eqn:thm1proofaux1} \\
&=\frac{1}{2}\log{\frac{1+\rho}{2\Delta e^{R_p}+\rho -1}} \label{eqn:thm1proofaux2},
\end{align}
where, (\ref{eqn:thm1proofaux1}) follows from $h(W) \leq \frac{1}{2} \log{(2\pi e \Var(W))}$ and (\ref{eqn:thm1proofaux2}) follows from $\Var(W)=1$ and $\Var(N)=\frac{2\Delta e^{R_p}+\rho -1}{1+\rho}$. Note that (\ref{eqn:thm1proofaux2}) corresponds to the upper bound in (\ref{eqn:partoneLUB}). A similar argument is used for the upper bound in (\ref{eqn:parttwoLUB}) where $W$ is a random vector of size two and we will skip the proof due to lack of space.

\section{Proof of Theorem \ref{thm:addGauss}} \label{app:proofaddGauss}

We compute $R_c(R_p)$ defined in (\ref{eqn:GWdef}), by improving the upper bounds in Theorem \ref{thm:main}. The lower bounds are the same as (\ref{eqn:parttwoLUB}) with the same validity region and (\ref{eqn:partoneLUB}) that is valid for $1-\rho \le \Delta e^{R_p} \le 1 -\sigma^2_{\theta}$. The upper bound of (\ref{eqn:addGaussp-one}) is derived below.

\subsection{Upper bound of (\ref{eqn:addGaussp-one}) - $W$ is random vector of size one}
The pair $(Z_x,Z_y)$ is jointly Gaussian that can be written as

\begin{align}
Z_X&=\sqrt{\alpha} V +N_X, \\
Z_Y&=\sqrt{\alpha} V +N_Y,
\end{align}
where $V\sim \mathcal{N}(0,1)$ is independent of the pair $(N_X,N_Y)$, 
\begin{align} \label{eqn:alpharegion}
0\leq \alpha \leq \rho -\sigma^2_{\theta},
\end{align}
and 
\begin{align}
K_{(N_X,N_Y)}=\begin{pmatrix}
1 -\sigma^2_{\theta} -\alpha & \rho-\sigma^2_{\theta} -\alpha \\
\rho -\sigma^2_{\theta} -\alpha & 1 -\sigma^2_{\theta} -\alpha
\end{pmatrix}.
\end{align}
Let $W=\theta +\sqrt{\alpha}V$, thus 
\begin{align}
X&=W+N_X \\
Y&=W+N_Y,
\end{align}
where $W$ is independent of the pair $(N_X,N_Y)$. Then we find that 
\begin{align}
&R_{X|W}(\Delta)+R_{Y|W}(\Delta) \nonumber \\
&=\frac{1}{2}\log\frac{\mathbb{E}[\Var{(X|W)}]\mathbb{E}[\Var{(Y|W)}]}{\Delta^2} \label{eqn:achievGrayWyner1} \\
&=\frac{1}{2}\log\frac{(\Var{(X)} - \Var{(\mathbb{E}[X|W])})(\Var{(Y)} - \Var{(\mathbb{E}[Y|W])})}{\Delta^2} \label{eqn:achievGrayWyner2}\\
&=\log\frac{1-\sigma^2_{\theta}-\alpha}{\Delta}=R_p, \label{eqn:achievGrayWyner3}
\end{align}
where (\ref{eqn:achievGrayWyner1}) follows because $X|W=w$ and $Y|W=w$ are Gaussian with same variance for any instance $w$, (\ref{eqn:achievGrayWyner2}) follows from the law of total variance and (\ref{eqn:achievGrayWyner3}) follows from $\mathbb{E}[X|W]=W$ and $\Var(W)=\sigma^2_{\theta}+\alpha$. Thus by using (\ref{eqn:achievGrayWyner3}) in (\ref{eqn:alpharegion}), the validity region is $1-\sigma^2_{\theta} \geq \Delta e^{R_p} \geq 1-\rho$. Let us now compute the upper bound, that is 
\begin{align}
R_c(R_p) \leq I(X,Y;W)&= h(X,Y)-h(N_X,N_Y) \\
&=\frac{1}{2}\log^+\frac{N^2(X,Y)}{(1-\rho) \left( 2 \Delta e^{R_p}+ \rho -1\right)},
\end{align}
that corresponds to Equation~\eqref{eqn:addGaussp-one}. A similar argument is used for Equation~\eqref{eqn:addGaussp-two} where $W$ is a random vector of size two and we will skip the proof due to lack of space.

\section*{Acknowledgment}
This work was supported in part by the Swiss National Science Foundation, postdoc mobility fellowship under Grant 199759
and under Grant 200364.

\bibliographystyle{IEEEtran}
\bibliography{GW}

\end{document}